\documentclass[cpp,a4paper,fleqn%
]{w-art}
\usepackage{times,cite,w-thm}
\theoremstyle{plain}

\theoremstyle{definition}

\usepackage[]{graphicx}

\usepackage{amssymb}
\usepackage{amsfonts}

\renewcommand{\vec}[1]{\mathbf{#1}}

\begin{document}
\DOIsuffix{theDOIsuffix}
\Volume{46}
\Month{01}
\Year{2007}
\pagespan{1}{}
\Receiveddate{XXXX}
\Reviseddate{XXXX}
\Accepteddate{XXXX}
\Dateposted{XXXX}


\title[Gradient correction and Bohm potential...]{Gradient correction and Bohm potential for 2D and 1D electron gases at a finite temperature}


\author[Moldabekov]{Zh. A.~Moldabekov\inst{1,2,3}\footnote{Corresponding author, email:~\textsf{zhandos@physics.kz}}}
\author[Bonitz]{M.~Bonitz\inst{1,}}

\author[Ramazanov]{T. S.~Ramazanov\inst{2,3}}

\address[\inst{1}]{Institut f\"ur Theoretische Physik und Astrophysik, Christian-Albrechts-Universit\"at zu Kiel,\\
 Leibnizstra{\ss}e 15, 24098 Kiel, Germany}

\address[\inst{2}]{Institute for Experimental and Theoretical Physics, Al-Farabi Kazakh National University,\\ 71 Al-Farabi str.,  
 050040 Almaty, Kazakhstan}
 
\address[\inst{3}]{Institute of Applied Sciences and IT, 40-48 Shashkin Str., 050038 Almaty, Kazakhstan}

\begin{abstract}
From the static polarization function of electrons in the random phase approximation  the quantum Bohm potential 
for the quantum hydrodynamic description of electrons, and the density gradient correction to the Thomas-Fermi free energy at a finite temperature
for the 2D and 1D cases are derived. The behavior of the Bohm potential and of the density gradient correction as a function of the degeneracy parameter is discussed.
Based on recent developments in the fluid description of quantum plasmas,  the Bohm potential for the high frequency domain is presented.  

\end{abstract}
\maketitle 

\section{Introduction}

A plasma is referred to as quantum plasma at densities and temperatures for which 
the electronic Fermi energy is larger or comparable to the thermal energy of a particle or when the characteristic de Broglie wavelength of adjacent particles is comparable with the mean distance between them.    
A comprehensive description of static and dynamic properties of quantum plasmas is challenging due to manifestation of the quantum non-locality, spin statistics, and exchange-correlation effects.  
Among other theoretical methods such as Green functions \cite{Schlunsen} or quantum Monte Carlo \cite{Schoof_CPP}, the density functional theory (DFT)  has become particularly popular due to the rapid development of the method over the last decades.
Particularly, the orbital-free formulation of the DFT allows for large scale simulation of quantum plasmas. The quality of the calculations based on the orbital-free density functional theory (OFDFT)  depends upon the implemented 
non-interacting free energy functional and exchange-correlation energy functional. It is well known that in many cases gradient corrections are required in addition to the local density approximation (LDA).
Since the works of Hohenberg and Kohn \cite{Kohn}, and Mermin \cite{Mermin2} on the density-functional theory, gradient corrections 
for the non-interacting free energy (the kinetic energy, in the case of the ground state) have been under consideration. 
In the two-dimensional case, the LDA was proved to be a very good approximation even when there are bound states \cite{Koivisto} or the system is inhomogeneous \cite{Eschrig}.
Additionally, the density gradient correction is related to the quantum Bohm potential in the quantum hydrodynamics \cite{michta_cpp15, Zhandos_QHD, Manfredi_1, Manfredi_2, Manfredi_3}.
 While most previous works focused a 3D system, much less is known  about the density gradient correction at a finite temperature for 2D and 1D systems. Fowler et al. \cite{Fowler} demonstrated for the first time that 
 the two-dimensional electron gas is realized in a n-type electron inversion layer.  
 The investigation of the electronic properties of III-V semiconductors led to the realization of high-mobility two-dimensional electronic systems at the interface of GaAs/AlGaAs and similar heterostructures.
Using modern electrostatic and etching techniques, the III-V semiconductor interface can be engineered to create low-dimensional electron systems including quantum wires and quantum dots \cite{Books}.

For the three dimensional case, the correct leading term of the gradient correction of the kinetic energy
was first obtained by Kirzhnitz \cite{Kirzhnitz}. He showed that the von Weizs\"acker gradient correction \cite{Weizsacker} must be 
corrected by an additional pre-factor $1/9$. The Feynman path-integral
method was used by Yang \cite{Yang} to obtain the single particle Green function and to prove the correctness of the factor $1/9$. 
For finite temperatures, the gradient correction of the non-interacting free energy was derived  by Perrot \cite{Perrot} considering the static ideal density response function of a uniform electron gas in the long wavelength limit. 

For the 2D and 1D cases, the leading term of the gradient correction was obtained by Holas et al. \cite{Holas} expanding the inverse ideal  polarization function of electrons.
The same result was obtained by Salasnich \cite{Salasnich} using the so-called  Kirzhnitz  expansion method.
It was revealed that, at the zero temperature limit, the leading term of the gradient correction to the kinetic energy is zero in the two dimensional case and negative in the one dimensional case.
Moreover,   Putaja et al.\cite{Putaja} have shown recently that in the zero temperature limit all higher order terms of the density gradient correction to the kinetic energy are equal to zero.
A non-zero leading term of the density gradient correction to the kinetic energy (in contrast to the free energy) for the 2D electron gas at a finite temperature in the semiclassical approximation 
was obtained by van Zyl et al \cite{Zyl1} 
using the semiclassical expansion of the diagonal Bloch density matrix.  
Additionally, attempts to go beyond the local density approximation, for the 2D system, were taken within so called average-density approximation \cite{Zyl}, 
and the gradient correction in terms of the effective  potential  energy \cite{Trappe} to mention but a few.

While the aforementioned works on  2D and 1D systems considered quantities related to the system in the ground state (like energy density functional),  
it is important that the Fermi energy in the systems of reduced dimensionality can be comparable with experimental temperatures \cite{Pouget_1, Pouget_2} meaning
that for the correct description of physical properties, such as screening and transport \cite{Sarma}, quantities describing the system at finite temperature (such as the free energy) must be considered. 
Therefore, in this work, the density gradient correction to the non-interacting free energy in the LDA and the finite temperature Bohm potential are considered on the basis of the polarization function in the random phase approximation (RPA) for an infinite system, i.e., neglecting possible effects due to a finite size of a system.
For the first time the density gradient correction to the non-interacting  free energy and the quantum Bohm potential \textit{at any temperature} within the quantum RPA in the 2D and 1D cases are given.

The paper is organized as follows: We first briefly discuss the connection of the first order density gradient correction and of
 the quantum Bohm potential with the expansion of the RPA polarization function in Sec.~\ref{s:HKM}.
In  Sec.~\ref{s:RPApol}, the analytical results for the expansion of the inverse polarization functions for the 2D and 1D cases as well as the discussion of the expansion convergence  are presented.
The finite temperature density gradient correction and the Bohm potential  are extracted from the RPA in Sec.~\ref{s:TradCor}. 
The result for the high frequency case is discussed in the conclusion.

\section{Relation between the polarization function in the RPA and the density gradient correction}\label{s:HKM}
The non-interacting free energy density  functional is expressed in terms of  the Tomas-Fermi free energy density functional  and the first order density gradient correction as follows \cite{Perrot}:
\begin{equation}\label{En}
F_{\rm id}([n],T)=\int f_0[n(\vec r)] \mathrm{d}\vec r+\frac{\hbar^2}{8m_e}\int \! \gamma_D \left(n, T\right) \frac{\mid \nabla n(\vec r)\mid^{2}}{n(\vec r)} \,\mathrm{d}\vec{r},
\end{equation}
where $f_0$ is the free energy density in the LDA, $D$ indicates the dimensionality, and  $\gamma_D$ is to be determined in this work.
Considering a small perturbation $n_1$ of the electron density around its mean value $n_0$, 
from the expansion of the inverse static ideal (RPA) polarization function  of a uniform electron gas, $\Pi_{\rm RPA}(\vec{k})^{-1}$,

\begin{equation}\label{K2}
\frac{1}{2\Pi_{\rm RPA}(\vec{k})}={\tilde a}_0
+{\tilde a}_2 \vec{k}^2 + {\tilde a}_4 \vec{k}^4+...,
\end{equation}
Perrot derived the following relations connecting $f_0$ and $\gamma_D$ with the expansion coefficients ${\tilde a_{0}}$ and ${\tilde a}_2$, respectively, \cite{Perrot}  

   \begin{align}
{\tilde a_{0}}[n_0]&=-\left. \frac{1}{2}\frac{\partial^2f_0[n]}{\partial n^2}\right|_{\rm n=n_0}, ~\quad {\tilde a}_2[n_0]=- \gamma_D(n_0, T) \frac{\hbar^2}{8m n_0}.
 \label{term_2}
\end{align}

The functional derivative of the leading term of the density gradient correction gives the Bohm potential of quantum hydrodynamics \cite{ZhandosPOP, Yan, Ciraci}:
\begin{multline}\label{Bohm_pot}
V_{B}= \frac{\hbar^2}{8m_e} ~\frac{\delta}{\delta n}\left(\int \! \gamma_D \left(n, T\right) \frac{\mid \nabla n(\vec r)\mid^{2}}{n(\vec r)} \,\mathrm{d}\vec{r}\right)
=-2 \frac{\hbar^2}{8m_e} \gamma_D \left(n, T\right) \frac{\vec{\nabla}^2 n}{n} +\mathcal{O}\Big(\left(n_1/n_0\right)^2\Big).
 \end{multline} 
 
 Below we consider the  static cases. The high frequency case is briefly discussed in the conclusion.
 
\section {Expansion of the inverse polarization function in low dimensions}\label{s:RPApol}

\subsection{2D electron gas}
For the two-dimensional case, we use the polarization function in the form given by Bret and Deutsch \cite{Bret}:
\begin{equation}\label{RPA2}
\Pi_{RPA}(k,\omega)=-\frac{k_F^2\chi_{0}^{2}}{2 \pi e^2 k}\left[g_2(u+z)-g_2(u-z)\right].
\end{equation}
where  $u=\omega/(kv_F)$, $z=k/(2k_F)$, $\chi_{0}^{2}=3/16 \left( \hbar\omega_p/E_F\right)^2=1/(\pi k_F a_B)$, $k_F=(3\pi^2n)^{1/3}$, $\omega_{p}^{2}=4\pi ne^2/m$,  $a_B=\hbar^2/me^2$ is the Bohr radius, 
$v_F$ is the Fermi velocity, and $k_F=(2\pi n)^{1/2}$.

The function $g_2$ in Eq.~(\ref{RPA2}) reads:
\begin{equation}\label{g2}
g_{2}(x)=2\pi\frac{x}{|x|}\int_0^x \! \frac{y\,\mathrm{d}y}{[\exp(y^2\theta^{-1}-\eta)+1]\sqrt{x^2-y^2}}-2i\pi\int_x^\infty \! \frac{y\,\mathrm{d}y}{[\exp(y^2\theta^{-1}-\eta)+1]\sqrt{y^2-x^2}},
\end{equation}
where $\theta=k_BT/E_{F}$ is the degeneracy parameter, $\eta=\mu/k_BT$ is the chemical potential which depends on the density according to the relation $\eta=\ln(\exp(\theta^{-1})-1)$, 
where the notation was introduced by Arista and Brandt for the dielectric function of the electrons gas in three dimensions \cite{Arista}.

In the static limit, $\omega=0$, Eq.~(\ref{RPA2}) can be written as:
\begin{equation}\label{RPAstat2}
\Pi_{RPA}(k)=-\frac{2k_F^2\chi_{0}^{2}}{e^2 k}\int_0^z \! \frac{y\mathrm{d}y}{[\exp(y^2\theta^{-1}-\eta)+1](z^2-y^2)^{1/2}},
\end{equation}
and simplifies in the long wavelength limit, $z\ll1$:
\begin{equation}\label{RPAstat2_1}
\Pi_{RPA}(k)\simeq-\frac{k_F\chi_{0}^{2}}{e^2}\frac{1}{[1+\exp(\frac{2z^2}{3\theta}-\eta)]}.
\end{equation}

For the 2D case, Eq.~(\ref{RPAstat2_1}) allows us to find the coefficients of the expansion Eq.~(\ref{K2}) of all orders:
\begin{align}
 {\tilde a}_0(n,T) &= - \frac{\pi e^2a_{B}}{2}\left[1+\exp(-\eta)\right], \quad 
{\tilde a}_{2l}(n,T) = -\frac{\pi e^2a_{B}}{2}\left(\frac{2}{3}\right)^l\frac{\exp(-\eta)}{l\,!\,\theta^{l}(2 k_F)^{2l}},
\label{eq:a2Di}
\end{align}
where $l=1,2,3,...$ is integer valued.

At finite temperature, ${\tilde a}_0$ in  Eq.~(\ref{eq:a2Di}) satisfies the relation (\ref{term_2}).
At $\theta\ll1$, making use of relation (\ref{term_2}) together with Eq.~(\ref{eq:a2Di}), correctly recovers the kinetic energy density in the LDA, $f_0[n]\to\tau[n]=\frac{\hbar^2\pi}{2m}n^2 $.

\subsection{1D electron gas}

Using the same notations as in the 2D case, the polarization function of the electron gas in one dimension can be expressed as 
\begin{equation}\label{RPA1}
\Pi_{RPA}(k,\omega)=-\frac{k_F\chi_0^2}{e^2 k}\left[g_1(u+z)-g_1(u-z)\right],
\end{equation}
where
\begin{equation}\label{g1}
g_{1}(x)=\int_{-\infty}^\infty \! \frac{\,\mathrm{d}y}{[\exp(y^2\theta^{-1}-\eta)+1](x+y)},
\end{equation}
and $k_F=\pi n/2$. In order to consider the chemical potential as a function of  $\theta$, the relation $2\theta^{-1/2}=I_{-1/2}(\eta)$ has to be inverted.

In the static limit, $\omega=0$, Eq.~(\ref{RPA1}) simplifies to 
\begin{equation}\label{RPAstat1}
\Pi_{RPA}(k)=\frac{k_F}{\pi E_F}\int_0^\infty \! \frac{\mathrm{d}y}{[\exp(y^2\theta^{-1}-\eta)+1](y^2-z^2)}.
\end{equation}
In the long wavelength limit, $z\ll1$, using the expansion $\frac{1}{y^2-z^2}\simeq\frac{1}{y^2}+\frac{z^2}{y^4}+...+\frac{z^{2j}}{y^{2j+2}}+..$, with $j=1,2,...$, 
and taking into account the definition of the Fermi integrals, one obtains the expansion 
\begin{equation}\label{1Dexp}
\Pi_{RPA}(k)\simeq\frac{k_F}{\pi E_F}\left(\frac{1}{2\theta^{1/2}}I_{-3/2}(\eta)+\frac{1}{8k_F^2\theta^{3/2}}I_{-5/2}(\eta)k^2+...\right).
\end{equation}
We find, using Eq.~(\ref{1Dexp}), that
\begin{equation}\label{K6}
\frac{1}{2\Pi_{\rm RPA}(k)}=\frac{\pi E_F\theta^{1/2} }{k_F I_{-3/2}(\eta)(1+\sum b_i z^{2i})},~\quad 
b_i\left(\left[n\right],\theta\right)= \frac{\theta^{-i}I_{-i-3/2}(\eta)}{I_{-3/2}(\eta)}.
\end{equation}

\begin{figure}[t]
 \centering
\includegraphics[width=0.33\textwidth]{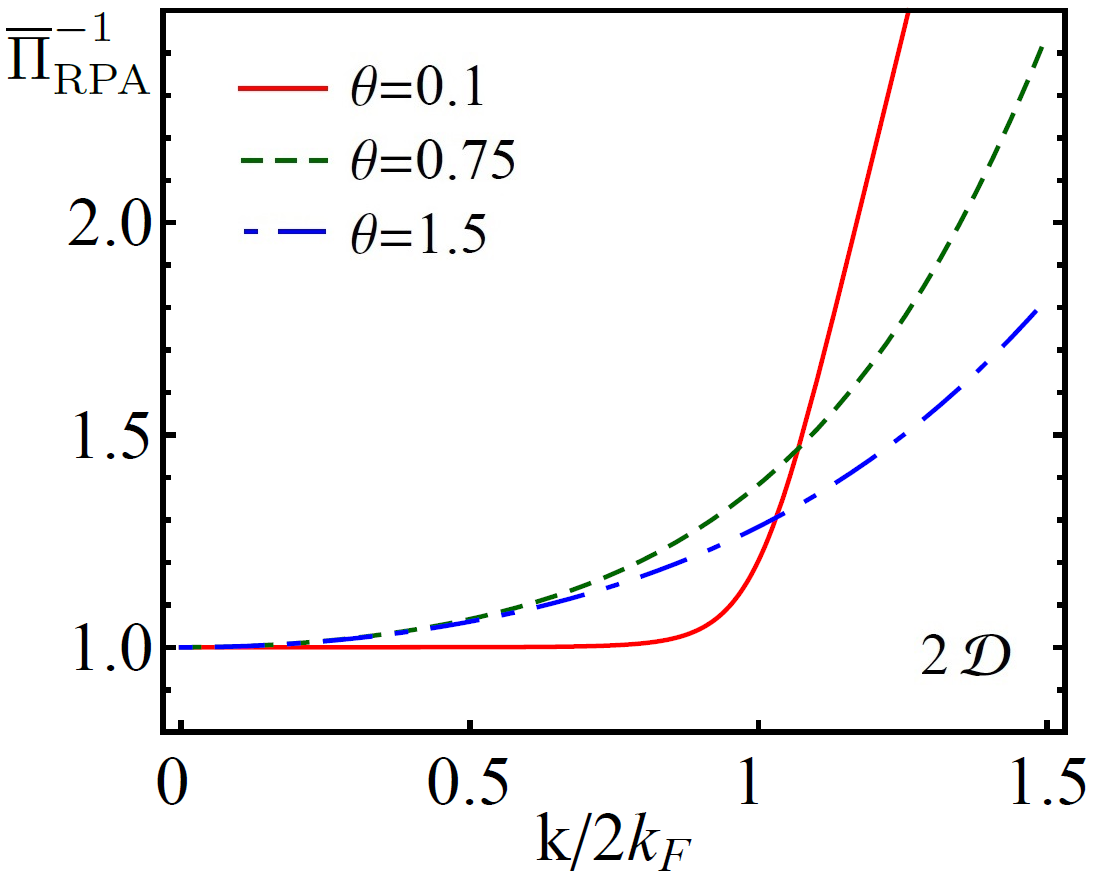}~\hspace{-3mm}\begin{scriptsize}a)\end{scriptsize}
\includegraphics[width=0.33\textwidth]{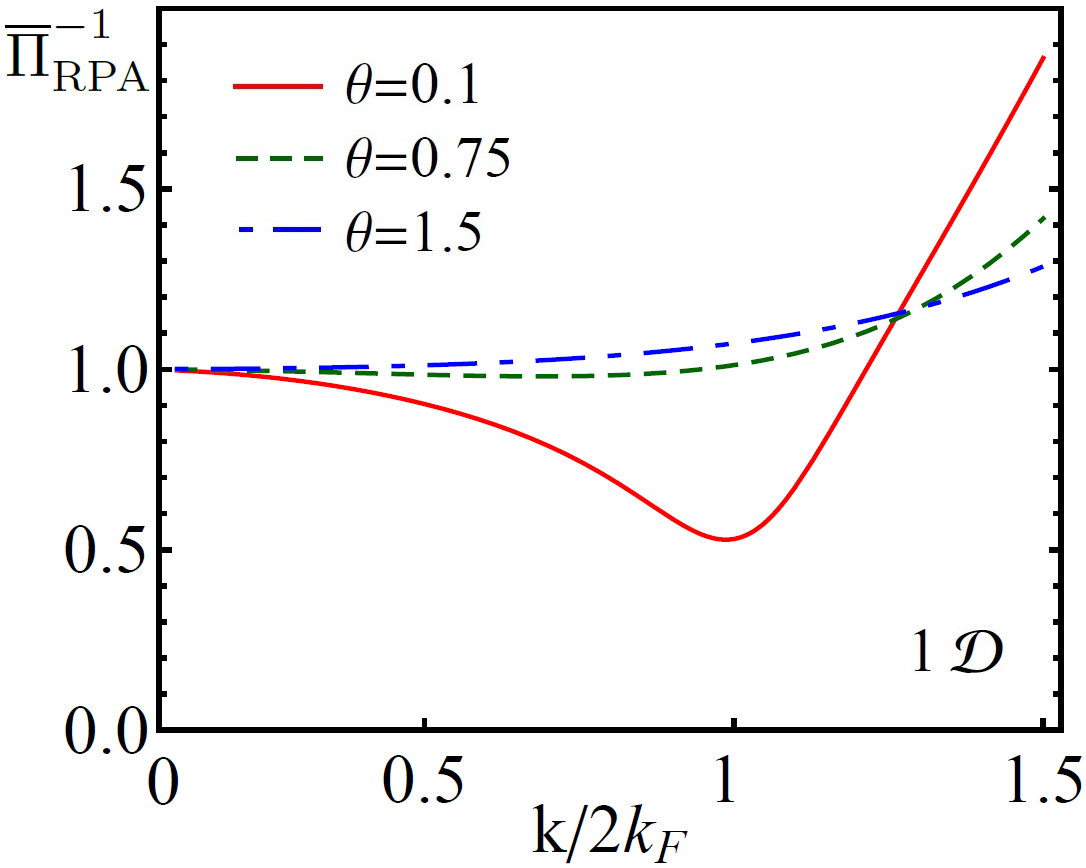}~\hspace{-3mm}\begin{scriptsize}b)\end{scriptsize}
\caption{The inverse static RPA polarization function in units of its value at $k=0$ for three values of the degeneracy parameter $\theta$: a) 2D electron gas and b) 1D electron gas.}
\label{fig:Kcurves}
\end{figure}

From Eq.~(\ref{K6}) we have obtained the first five non-vanishing coefficients of the expansion, Eq.~(\ref{K2}), for the 1D case:
\begin{align}
 {\tilde a}_0(n,T) &= \frac{\pi E_{F}\theta^{1/2}}{k_F I_{-3/2}(\eta)},~\quad 
 \frac{{\tilde a}_2(n,T)}{{\tilde a}_0(n,T)} = -\frac{b_1}{4 k_F^2},~\quad 
 \frac{{\tilde a}_4(n,T)}{{\tilde a}_0(n,T)} =\frac{b_1^2-b_2}{16 k_F^4},\quad 
 \label{eq:a1d2}
 \\
 \frac{{\tilde a}_6(n,T)}{{\tilde a}_0(n,T)} &=\frac{-b_1^3+2b_1b_2-b_3}{64 k_F^6},~\quad 
 \frac{{\tilde a}_8(n,T)}{{\tilde a}_0(n,T)} = \frac{b_1^4-3b_1^2b_2+b_2^2+2b_1b_3-b_4}{256 k_F^8}.
 \label{eq:a1d8}
\end{align}

It is checked that the coefficient ${\tilde a}_0$, Eq.~(\ref{eq:a1d2}), satisfies the relation (\ref{term_2}).

In Fig.~\ref{fig:Kcurves}, the inverse static polarization function in units of its value at $k=0$,  $\overline{\Pi}_{\rm RPA}^{-1}(k)= \Pi_{\rm RPA}(0)/\Pi_{\rm RPA}(k)$, is calculated using Eqs. (\ref{RPA1}), and (\ref{RPA2}) for different values of the degeneracy
parameter. 
At $k/2k_F> 1$, $\overline{\Pi}_{\rm RPA}^{-1}$ decreases with increase in $\theta$. 
In contrast, in the long wavelength limit, $k/2k_F<1$, the dependence of  $\overline{\Pi}_{\rm RPA}^{-1}$
on the degeneracy parameter is nonmonotonic.

\begin{figure}[t]
 \centering
\includegraphics[width=0.37\textwidth]{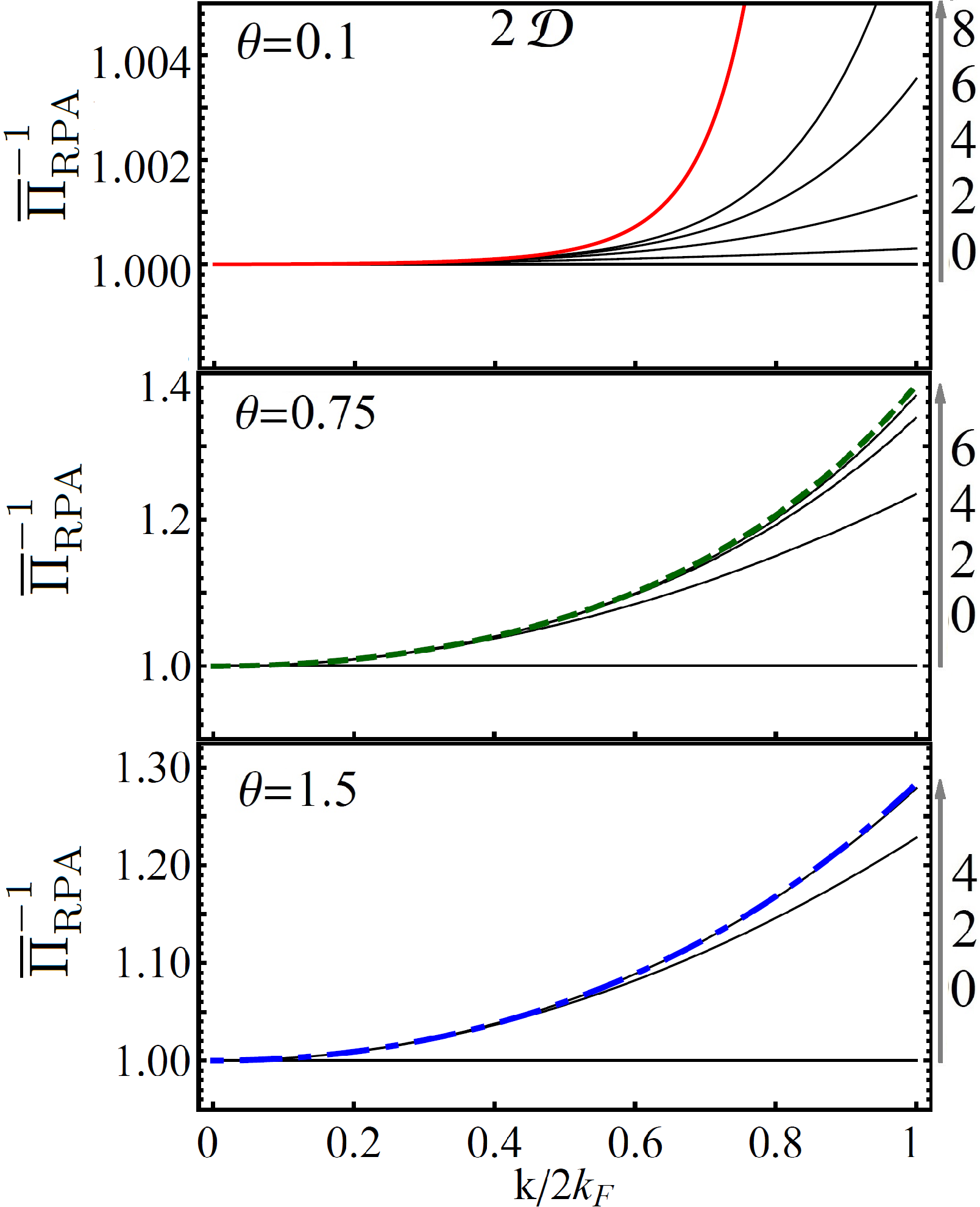}~\hspace{-2mm}\begin{scriptsize}a)\end{scriptsize}
\includegraphics[width=0.356\textwidth]{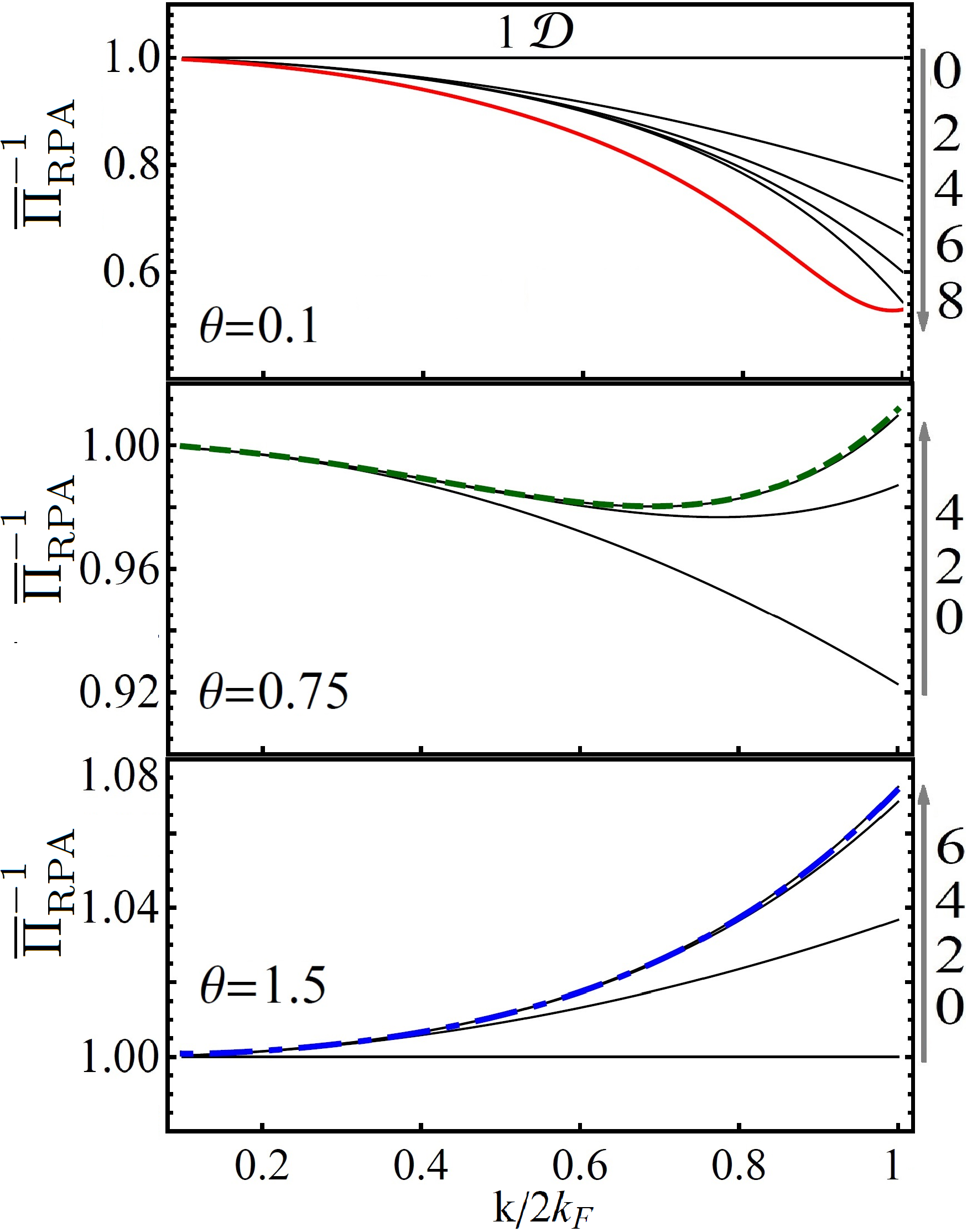}~\hspace{-2mm}\begin{scriptsize}b)\end{scriptsize}
\caption{(Color online)  The convergence of the expansion (\ref{K2}) is illustrated for the values of the degeneracy parameter 0.1, 0.75, and 1.5: a) 2D electron gas and b) 1D electron gas. 
Solid thin (black) curves correspond  to the different maximal orders of the expansion that are indicated on the right hand side of the figure. 
At $\theta=0.75$ and $\theta=1.5$, curves corresponding to the higher order corrections are indistinguishable from the exact curves. 
For the 2D case, the coefficients of the expansion are given  in Eq.~(\ref{eq:a2Di}).
 For the 1D case, the coefficients of the expansion of $1/(2\Pi_{RPA}(k))$,  Eq.~(\ref{K2}),  are given in Eqs.~(\ref{eq:a1d2}) and (\ref{eq:a1d8}).}
\label{fig:K2D}
\end{figure}

In Fig.~\ref{fig:K2D}, the convergence of the expansion (\ref{K2}) is demonstrated for different $\theta$.
In the 2D case, in the limit of low temperatures, $\theta\simeq 0.1$ , the zero order correction, $\Pi_{\rm RPA}^{-1}(0)=2\tilde a_0$
(\ref{eq:a2Di}), gives nearly exact result up to $k/2k_F\simeq 0.4$ as it is seen from Fig.~\ref{fig:K2D}. However, the  convergence of the expansion (\ref{1Dexp}) at $k\gtrsim k_F$ is rather bad. 
At higher temperatures ($\theta=0.75$ and $\theta=1.5$), taking into account of the first non-vanishing correction, $\tilde 2a_2 (k/2k_F)^2$ (see Eq.~(\ref{eq:a2Di})), is enough for the description
of $\Pi_{\rm RPA}^{-1}(k)$ at $k \ll k_F$.

For the 1D case, the convergence of the expansion, Eq.~(\ref{K2}), at different values of $\theta$  is demonstrated in Fig.~\ref{fig:K2D}.b.
At the low temperature limit ($\theta=0.1$), the convergence of the expansion (\ref{K2}),  with the coefficients Eqs.~({\ref{eq:a1d2})-(\ref{eq:a1d8}}), is slower in comparison with both the 3D and 2D cases, whereas at 
high temperatures ($\theta=0.75$ and $\theta=1.5$),  the convergence of the expansion (\ref{K2}) is more rapid as in the case of higher dimensionality.   

%
%
%
%

\section {Gradient correction and Bohm potential}\label{s:TradCor}

From the expansion of  the inverse value of the polarization functions in the long wavelength limit, in the static case, $\gamma_D$ required for the density gradient correction in Eq.~(\ref{En}) and for the Bohm potential in Eq.~(\ref{Bohm_pot}) is determined 
 using the relation (\ref{term_2}). For the two dimensional case, $D=2$, we have: 
\begin{equation}\label{grad_cor2D}
\gamma_2\left[n\right] =\frac{1}{3}\frac{\theta^{-1}}{\exp(\theta^{-1})-1},
\end{equation}
and for the one dimensional case $\gamma_1$ reads:
\begin{equation}\label{grad_cor1D}
\gamma_1\left[n\right] =\frac{1}{3}2\theta^{-1/2}\frac{I_{-1/2}^{\prime\prime}(\eta)}{I_{-1/2}^{\prime 2}(\eta)},
\end{equation}
 where the derivatives of $I_{-1/2}$ are taken with respect to $\eta$.

In the case of three dimensions, $\gamma_3$ has the following form \cite{Kirzhnitz, Perrot, ZhandosPOP}:
\begin{equation}\label{grad_cor3D}
\gamma_3\left[n\right] =\frac{1}{9}4\theta^{-3/2}\frac{I_{-1/2}^{\prime}(\eta)}{I_{-1/2}^{2}(\eta)}.
\end{equation}

The pre-factor $\gamma_D$ in the limit of high temperatures has the value equal to $1/3$ for all three cases, D=1, 2, 3.
In the 3D case, this result is well known since the work of Kirzhnitz in 1957 \cite{Kirzhnitz}. In the 2D case, at $\theta\gg1$, using the relation $f_0(n.T)=\tau_0(n,T)-T\sigma(n,T)$ between the free energy density, the kinetic energy density,
and so-called entropic contribution ($\sigma=-\left.\frac{\partial f_0}{\partial n}\right|_T$) in combination with Eqs.~(\ref{grad_cor2D}) and (\ref{term_2}) we recover the result obtained recently by van Zyl et al.\cite{Zyl1} for the density gradient correction to the kinetic energy.

From Eq.~(\ref{grad_cor2D}), it is clear that in the strongly degenerate limit $\theta\ll1$, the leading term of the gradient correction for the 2D case tends to zero as $\sim\exp(-\theta^{-1})$.
In the one dimensional case, in the limit $\theta\ll1$, the pre-factor of the leading term of the gradient correction Eq.~(\ref{grad_cor1D}) becomes negative, $\gamma_1=-1/3$, in agreement with the 
results of Salasnich \cite{Salasnich} and Holas et al.\cite{Holas}. In the 3D case, at $\theta\ll1$, the value of the $\gamma_D$ equals to $1/9$ as it was first obtained by Kirzhnitz \cite{Kirzhnitz}.

\begin{figure}[h]
\centering
\includegraphics[width=0.5\linewidth]{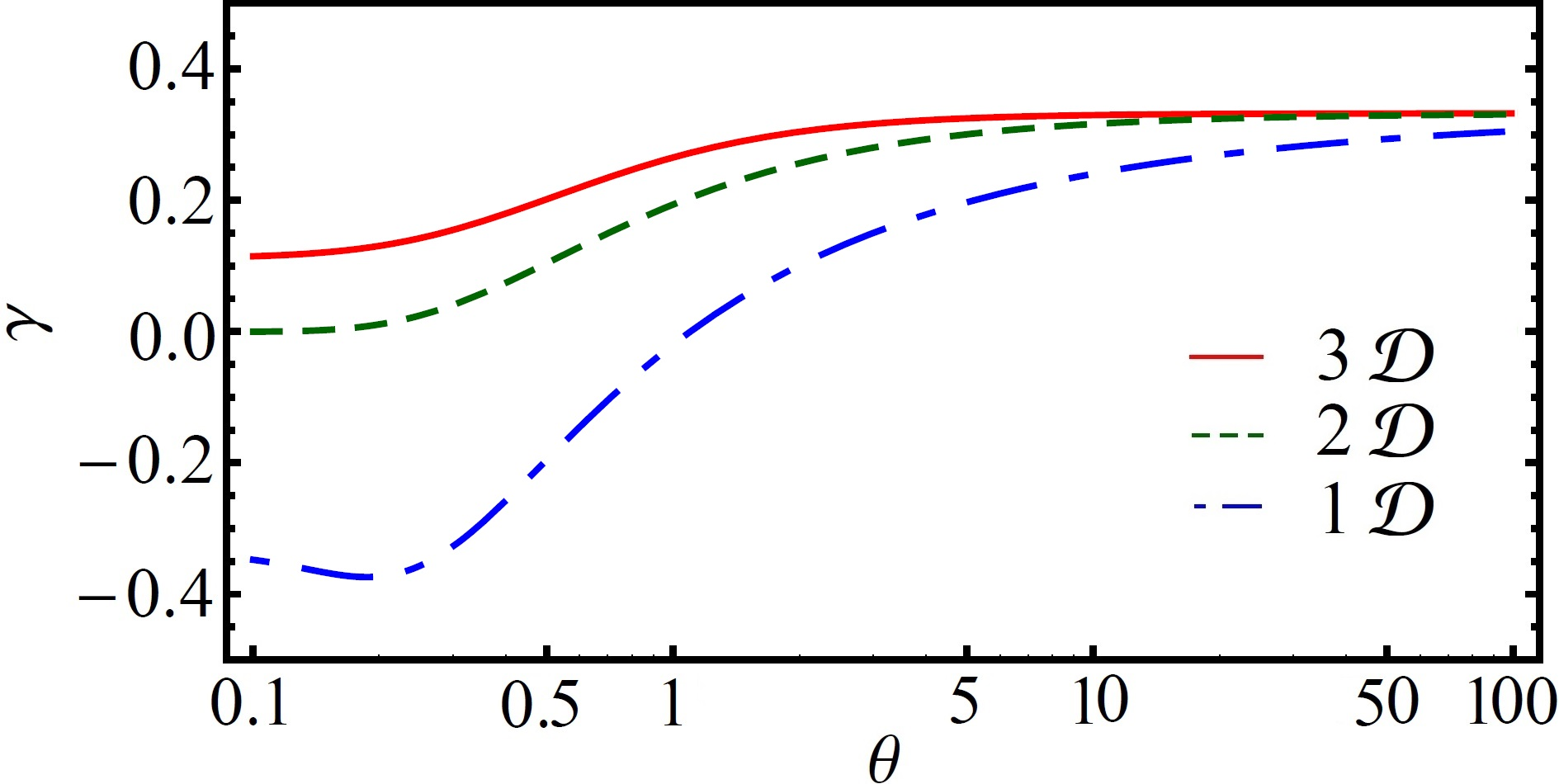}
\caption{(Color online) The dependence of the quantum Bohm potential's prefactor $\gamma_D$ in the static case, Eq.~(\ref{Bohm_pot}),  
on  $\theta$ is shown for the 3D, 2D, and 1D cases. It is seen that at the high temperature limit $\gamma_D$ tends to $1/3$.
 In the strongly degenerate limit $\theta\ll1$: $\gamma_1=-1/3$, $\gamma_2=0$, and $\gamma_3=1/9$.}
\label{fig:f1}
\end{figure}

The value of the $\gamma_D$ explicitly depends only on $\theta$, which in turn, at given temperature, depends on the local value of the density.
In the Fig.~\ref{fig:f1} the dependence of $\gamma_D$ on $\theta$ is shown.  Fig.~\ref{fig:f1}  confirms the analytical features which we have mentioned. 
Additionally, it is seen that the parameter $\gamma_1$ changes sign at $\theta\simeq 1.1$.
It reflects the fact that, in the 1D case, $\Pi_{\rm RPA}^{-1}(k)$ changes the sign of its slope from the negative to  the positive value at  $k/k_F\ll1$ (see Figs.~\ref{fig:Kcurves}.b and \ref{fig:K2D}.b). 
Another distinctive feature of the pre-factor $\gamma_1$ for the one-dimensional case from its analogues at $D=2$ and $D=3$ is that $\gamma_1$ has its minimum $\gamma_1(\theta_{\rm min})\simeq-0.374$ at $\theta_{\rm min}\simeq 0.19$, 
not at $\theta=0$ as $\gamma_2$ and $\gamma_3$.  

\section{Conclusion}
The finite temperature density gradient correction to the free energy in the LDA for the 2D and 1D cases is derived. 
It was shown that in the high temperature limit  the pre-factor $\gamma_D$ of the density  gradient correction is equal to $1/3$ exactly as in the 3D case.
 Our results for the density gradient correction are in agreement with the results of the previous works
 for the zero temperature case \cite{Salasnich, Holas} and for the high temperature limit \cite{Zyl1}.
The relation (\ref{term_2}), linking the non-interacting free energy density with the potentials for the quantum hydrodynamics, 
allowed us to determine the Bohm potential at a finite temperature in the static case. This, for instance, allows one to correctly calculate a statically screened potential of an ion (or impurity) \cite{ZhandosPOP}, 
but not in the case of dynamical screening (when the wake effect is important) \cite{Zhandos_CPP}. 
Presented results are also applicable in the low frequency regime, i.e., $\omega<kv_F$, but not in the high frequency case ($\omega\gg \hbar^2k_F^2/2m$).
The high frequency regime is of interest for the consideration of electron plasmons on the basis of quantum hydrodynamics.
For this case,  Perrot's relations (Eqs.~(\ref{term_2}), and (\ref{term_2})) were extended to the dynamic (high frequency) domain in Ref.~\cite{Zhandos_QHD}. 
Therefore, by analyzing the polarization functions at $\omega\gg \hbar^2k_F^2/2m$ and $k\ll2k_F$ we revealed that in the 2D case $\gamma_2=0$, and in the 1D case $\gamma_1=1$. 
 In Refs.~\cite{Akbari, Zhang}  plasmons in 2D quantum systems at $\theta\ll1$ were studied on the basis of the QHD model taking into account 
the non-zero Bohm potential with the pre-factor $\gamma$ equal to one. In this regard, it is instructive to note that the Bohm potential is not a universal 
potential describing the non-locality for arbitrary geometry (parameters) and, therefore, using Eq.~(\ref{term_2}) one always must 
check whether  $\gamma\neq 0$ for the present physical problem. Of course, the fact that $\gamma_2=0$,  in the high frequency regime 
and in the static ground state ($\omega=0$, $\theta\to0$) of the 2D system,  does not mean that there is no quantum non-locality at all  ~\cite{Trappe, Zyl}. 
For the 2D system in the static case, van Zyl et.al. \cite{Zyl} consistently introduced the non-vanishing density gradient correction applying the average-density approximation, which goes well beyond the LDA, and
 Trappe et.al. \cite{Trappe} derived the gradient correction in terms of the effective potential for the DFT formulated as a joint functional of both the single-particle density and the effective potential
energy \cite{Englert}. In Ref.~\cite{Zhandos_QHD}, the relation between the second order functional derivative of the non-interacting 
free energy functional and the dynamic polarization function in the RPA was established in the framework of the QHD model. This result provides the basis for the construction of 
non-local potentials for the fluid description of the electrons in the high frequency regime.

For reference we note that, at $\omega\gg \hbar^2k_F^2/2m$, we have found ${\tilde a_0}=-\frac{3\pi \hbar^2}{2m}\theta^2I_{1}(\eta)$ for the 2D case, and in the 1D case ${\tilde a_0}=-\frac{\pi^2\hbar^2}{8m}n$. 
Making use of these coefficients in the quantum hydrodynamics correctly reproduces the plasmons dispersion relation at $k/k_F \ll 1$ (for more details see Ref.~{\cite{Zhandos_QHD}}). 
The results of this work provide a better understanding of the non-interacting free energy density functional of the electrons and their applicability for low-dimensional systems in the QHD.
At the same time, in low-dimensional systems, correlation effects are particularly strong and their importance for the description by the QHD model  should be verified. This is can be done on the basis of other  methods such as Green functions \cite{Schlunsen}, 
quantum Monte Carlo \cite{Schoof_CPP} or an extended fluid model \cite{Hanno_CPP}.

\subsection*{Acknowledgments}
Zh.A. Moldabekov gratefully acknowledges funding from
the German Academic Exchange Service (DAAD) under program number 57214224.
   This work has been supported by the Ministry of Education and Science of Kazakhstan under Grant No. 0263/PSF (2017).

\end{document}